\documentclass[aps,twocolumn,prb,showpacs,superscriptaddress,floatfix]{revtex4-1}

\usepackage{amsfonts,amsmath,graphicx}

\newcommand{\ket}[1]{|#1\rangle}
\newcommand{\moy}[1]{\langle #1\rangle}

\begin{document}

\title{Quantum dynamics of a driven three-level Josephson-atom maser}
\author{N.~Didier}
\email{Nicolas.Didier@sns.it}
\affiliation{NEST, Scuola Normale Superiore and Istituto di Nanoscienze - CNR, Pisa, Italy}
\author{Ya.~M.~Blanter}
\email{Y.M.Blanter@tudelft.nl}
\affiliation{Kavli Institute of Nanoscience, Delft University of Technology, Lorentzweg 1, 2628 CJ Delft, The Netherlands}
\author{F.~W.~J.~Hekking}
\email{Frank.Hekking@grenoble.cnrs.fr}
\affiliation{Laboratoire de Physique et de Mod\'elisation des Milieux Condens\'es, Universit\'e Joseph Fourier and CNRS, BP 166, 38042 Grenoble, France}

\begin{abstract}
Recently, a lasing effect has been observed in a superconducting nano-circuit where a Cooper pair box, acting as an artificial three-level atom, was coupled to a resonator~\cite{Astafiev}. 
Motivated by this experiment, we analyze the quantum dynamics of a three-level atom coupled to a quantum-mechanical resonator in the presence of a driving on the cavity within the framework of the Lindblad master equation. 
As a result, we have access to the dynamics of the atomic level populations and the photon number in the cavity as well as to the output spectrum. 
The results of our quantum approach agree with the experimental findings. 
The presence of a fluctuator in the circuit is also analyzed.
Finally, we compare our results with those obtained within a semiclassical approximation.
\end{abstract}
\maketitle

\section{Introduction}

In recent years, a considerable progress has been achieved in the field of quantum manipulation with nano-circuits based on the Josephson effect~\cite{reviews,NazarovBlanter}. 
This progress has been initially inspired by the ideas of quantum information processing~\cite{Nielsen}, which require the physical realization of qubits as an elementary quantum information unit. 
The degree of control achieved in Josephson-based qubits is so high that these systems have become a test-bed for the ideas of quantum mechanics such as quantum noise detection~\cite{SchoelkopfNazarov,Astafiev1,mariantoni}, quantum measurements~\cite{Mooij,Martinis}, and realization of circuit-quantum electrodynamics (QED)~\cite{Wallraff,Schoelkopf}. 
Very recently, the experimental realization of a single Josephson-atom laser has been reported~\cite{Astafiev}.

In the experiment of Ref.~\onlinecite{Astafiev}, a Cooper pair box (CPB) is coupled to a superconducting waveguide resonator. The CPB is used as a three-level artificial atom. 
While the lowest two levels constitute a qubit, population inversion is achieved with the Josephson quasiparticle (JQP) cycle~\cite{JQP} involving the third level. 
The lasing condition can be determined from the steady state photon number, pointing out that a too strong pumping suppresses the lasing action~\cite{ashhab}.
Experimentally, evidence for lasing action was found through measurements of the output power spectrum of the resonator. 
An additional driving was applied on the cavity to induce phase locking, thereby enhancing the lasing effect. 
The results are consistent with theoretical work on a two-level atom coupled to a resonator~\cite{brosco, andre}. 
Studies concerning the coupling between a superconducting qutrit and a resonator have been also carried out~\cite{qutrit}.
However, currently no quantitative results are available for the dynamics of a three-level Josephson atom coupled to a resonator.

In this paper we present a theoretical analysis of the lasing effect observed in the experiment~\cite{Astafiev}, using a quantum-mechanical approach. 
We first obtain the complete time evolution of the system, including transient effects upon switching on the pumping. 
We obtain estimates for the characteristic time scales of the corresponding dynamics. 
Then, we study the output spectrum of the cavity field, with and without an additional driving applied to the cavity. 
We show that the latter requires a full quantum treatment and cannot be obtained from a semiclassical approximation. 
Our results are in good agreement with the experimental findings. 
The possibility to induce off-resonance lasing, observed in the experiment with a second hot spot, is also considered by adding a two level fluctuator in the circuit. 
Since our model is system-independent it can be applied to the study of other circuit-QED implementations, such as the currently much-studied transmon~\cite{transmon}.

\section{Model}

   \begin{figure}[b]
   \centering
   \includegraphics[width=\columnwidth]{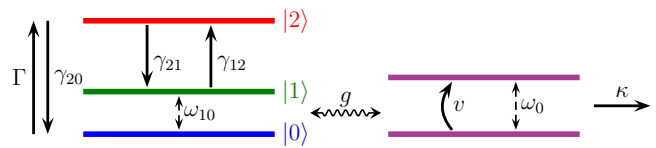}
   \caption{Scheme of the three-level system coupled to a cavity with the corresponding transition rates.}
   \label{figscheme}
   \end{figure}

The system under consideration is depicted in Fig.~\ref{figscheme}. 
It is composed of the three-level artificial atom $\{|0\rangle,|1\rangle,|2\rangle\}$, with an energy difference $\hbar\omega_{10}$ between the ground state and the first excited state, and the cavity with a mode frequency $\omega_0/2\pi$. 
These two subsystems are coupled coherently according to the Jaynes-Cummings Hamiltonian
   \begin{equation}
   H=\tfrac{1}{2}\hbar\omega_{10}\sigma_z
   +\hbar\omega_0 a^\dag a
   +i\hbar g(\sigma_{01}a^\dag-\sigma_{10}a),
   \label{Heigen}
   \end{equation}
where $\sigma_{ij}=|i\rangle\langle j|$, $\sigma_z=\sigma_{11}-\sigma_{00}$ and $a$ ($a^\dag$) is the canonical bosonic annihilation (creation) operator of a photon in the cavity. 
The dynamics of the third level $\ket{2}$ is described with a Lindbladian, as presented below.
This Hamiltonian is obtained after applying the rotating wave approximation (RWA), valid when the coupling strength is small compared to the typical frequency of the isolated subsystems, which will be the case in the following. 
The legitimacy of the RWA is also established through numerical simulations.
The atom is pumped from $|0\rangle$ to $|2\rangle$ at the rate $\Gamma$, the state $|2\rangle$ decays to $|1\rangle$ at the rate $\gamma_{21}$. 
The reverse processes occur at the rates, respectively, $\gamma_{20}$ and $\gamma_{12}$. 
Finally, the cavity has a damping rate $\kappa$, conferring to the photons a lifetime $\kappa^{-1}$. 
As a consequence, the time evolution of the total density matrix $\rho$ has two contributions: 
the evolution due to the coherent coupling between the artificial atom and the cavity according to Hamiltonian~\eqref{Heigen} and 
the evolution due to the incoherent processes and controlled by the Lindbladian
$L=L_\Gamma+L_{\gamma_{21}}+L_{\gamma_{20}}+L_{\gamma_{12}}+L_\kappa$ with
   \begin{align}
   L_{\gamma_{ij}}\rho&=\tfrac{1}{2}\gamma_{ij}(2\sigma_{ji}\rho\sigma_{ij}-\sigma_{ii}\rho-\rho\sigma_{ii}),\\
\intertext{for the pumping and the relaxation rates, noting $\Gamma\equiv\gamma_{02}$, and}
   L_\kappa\rho&=\tfrac{1}{2}\kappa(2a\rho a^\dag-a^\dag a\rho-\rho a^\dag a),
   \end{align}
for the damping of the cavity.
   \begin{figure}[t]
   \centering
   \includegraphics[height=6.0cm]{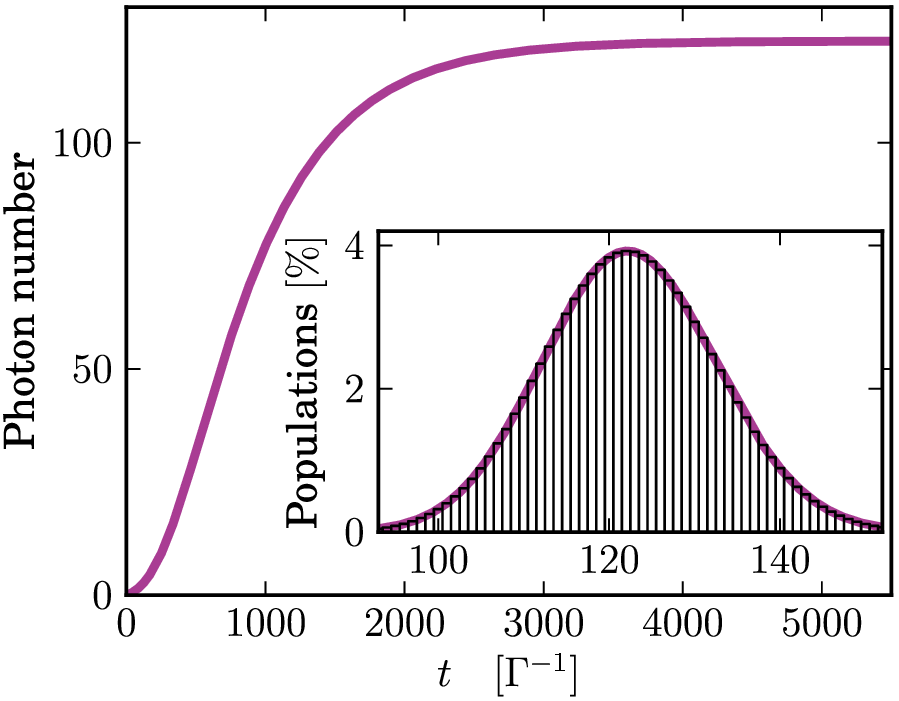}
   \caption{(Color online) Time evolution of the photon number in the cavity. 
   The coupling and various rates correspond to the experimental parameters at the resonance, as discussed in the text.
   Inset: Distribution of the photon population in the steady state (histogram) compared to the corresponding binomial distribution (solid line).}
   \label{figtransphot}
   \end{figure}
These expressions are obtained in the Born-Markov approximation, supposing a weak coupling between the system and the environment. 
The resulting time evolution of the density matrix satisfies the master equation~\cite{refsL}
   \begin{equation}
   \dot{\rho}(t) = \frac{1}{i\hbar}[H,\rho(t)] + L\rho(t).
   \label{lindbladian}
   \end{equation}
To characterize the coherence properties of the emitted field we calculate the output spectrum $\widehat{S}(\omega)$ defined as the Fourier transform of the cavity correlator
   \begin{equation}
   \widehat{S}(\omega)=\lim_{t\to\infty}\int_{-\infty}^{+\infty}\!\mathrm{d}\tau e^{-i\omega\tau}\moy{a^\dag(t+\tau)a(t)}.
   \label{spectrum}
   \end{equation}
The output spectrum is obtained from the steady-state density matrix using the quantum regression theorem~\cite{QRT}, valid within our approach, which establishes a matrix relation between them. 
We also include the possibility to drive the cavity with the additional pumping
   \begin{equation}
   H_d=i\hbar v(e^{-i\varpi t}a^\dag-e^{i\varpi t}a),
   \label{Hdrive}
   \end{equation}
where $\varpi$ is the detuning from the cavity frequency. 
The amplitude can be expressed in terms of the photon number $N_0$ created by the driving~\cite{Haroche} $v=\kappa\sqrt{N_0}/2$. 
If the emitted field is coherent, the injection locking effect occurs and the cavity field oscillates at the same frequency as the driving field $\omega_0+\varpi$.

   \begin{figure}[t]
   \centering
   \includegraphics[height=6.0cm]{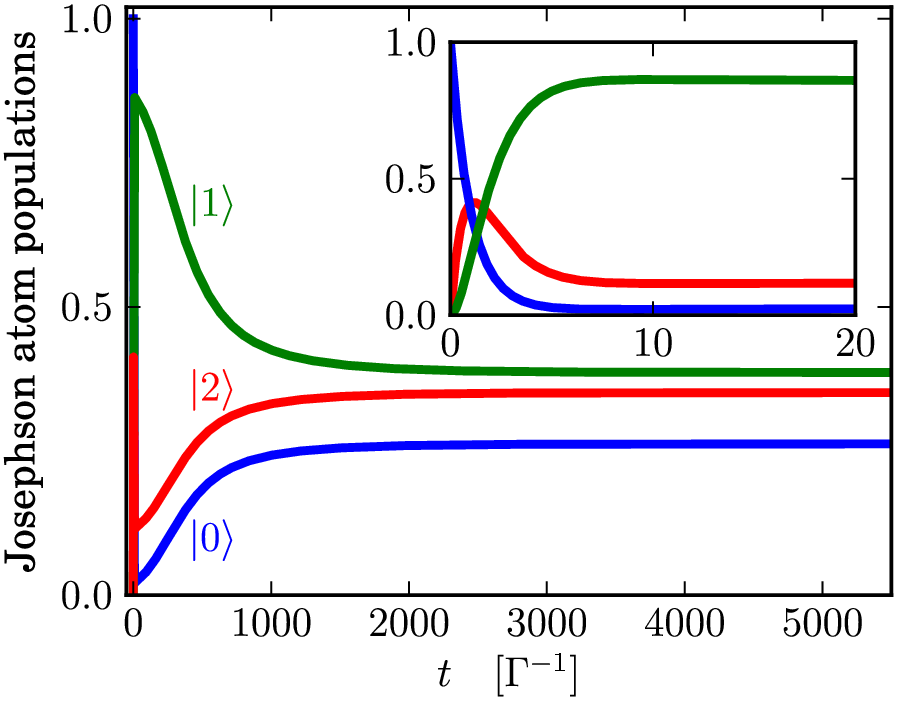}
   \caption{(Color online) Time evolution of the level populations at the resonance (level $|0\rangle$ in blue, $|1\rangle$ in green, and $|2\rangle$ in red).
   The dynamics of the first pump cycles for a qubit initially in the state $\ket{0}$ is presented in the inset.}
   \label{figtransqubit}
   \end{figure}

\section{Choice of parameters}

The results that we present below have been obtained using a particular choice for the numerical values of the various system parameters, corresponding to those of the experiment~\cite{Astafiev}. 
The three-level atom is a CPB, the properties of which are controlled by an external parameter, the dimensionless gate-voltage $n_g$. 
Varying the parameter $n_g$ corresponds to rotating the charge basis around the state $\ket{2}$ by an angle $\theta$ defined by
$\tan2\theta=E_J/[E_C(n_g-1)]$, 
where $E_J$ and $E_C$ are the Josephson energy and the charging energy of the CPB, respectively ($E_C/E_J\simeq15$).
The qubit energy then reads $\hbar\omega_{10}=E_J/\sin2\theta$, 
the coupling varies like $\sin2\theta$, 
the rates $\Gamma$ and $\gamma_{21}$ are proportional to $\cos^2\theta$, 
and the rates $\gamma_{20}$ and $\gamma_{12}$ are proportional to $\sin^2\theta$.
In the experiment, the cavity frequency is $\omega_0/2\pi\simeq10\,\mathrm{GHz}$; the resonance condition $\omega_{10} = \omega_0$ for the lowest two atom levels and the cavity is achieved when the parameter $n_g =1.1$.
At this working point, the atom-cavity coupling frequency is $\bar{g}/2\pi=44\,\mathrm{MHz}$. 
Population inversion is achieved using the JQP cycle; the relevant rates are $\Gamma=4.2\,\mathrm{GHz}$, $\gamma_{21}=3.3\,\mathrm{GHz}$, $\gamma_{20}=0.29\,\mathrm{GHz}$, and $\gamma_{12}=0.37\,\mathrm{GHz}$.
The damping rate is set to $\kappa=8.2\,\mathrm{MHz}$.
What can be measured experimentally is the spectrum of the cavity, Eq.~\eqref{spectrum}, with or without an additional driving Eq.~\eqref{Hdrive}. 
The photons being emitted at an energy of $10\,\mathrm{GHz}$, this lasing effect is actually a masing effect.

   \begin{figure}[t]
   \includegraphics[height=6.0cm]{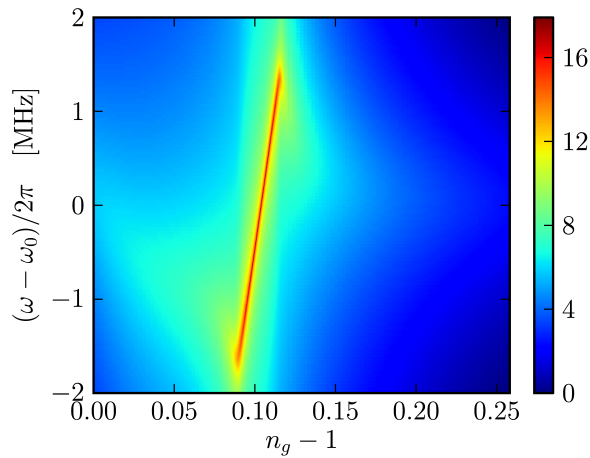}
   \caption{(Color online) Density plot of the spectrum (logarithmic scale) from our fully-quantum model as a function of the probing frequency and the reduced gate charge.}
   \label{fig2Dspectrum}
   \end{figure}

\section{Discussion of the results}

\begin{figure}[t]
   \centering
   \includegraphics[height=6.0cm]{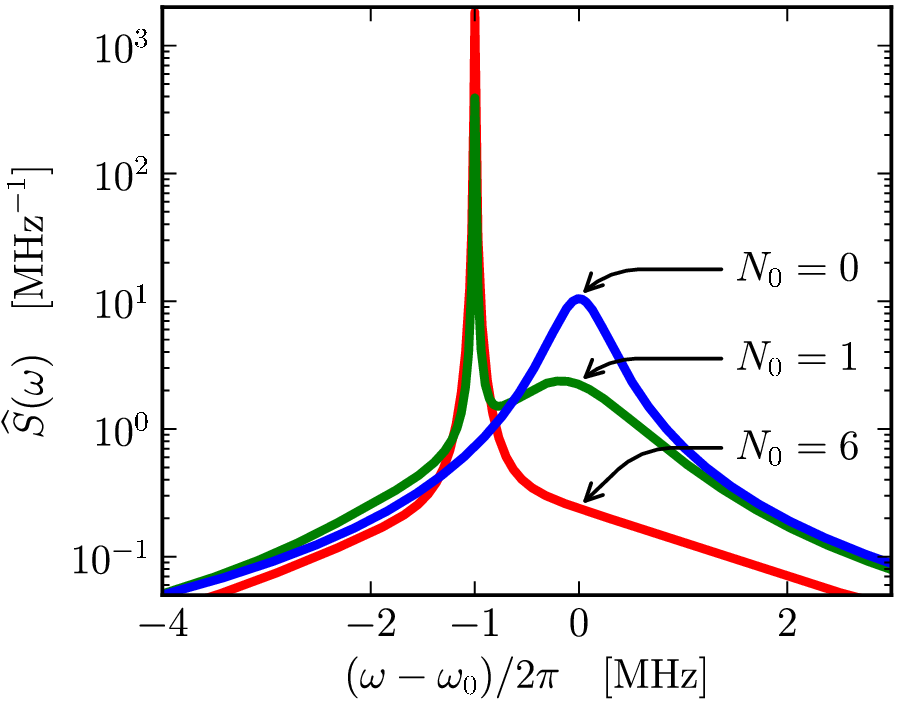}
   \caption{(Color online) Spectrum in the presence of an additional driving on the cavity ($\varpi/2\pi=-1\,\mathrm{MHz}$).
   The damping $\kappa$ has been increased fivefold.}
   \label{fig2Dspectrumdrive}
   \end{figure}

The Lindblad master equation Eq.~\eqref{lindbladian} gives access to the time evolution of the photon number in the cavity and of the Josephson atom level populations. 
The dynamics of these quantities is shown in Figs.~\ref{figtransphot} and~\ref{figtransqubit}, using the parameters given above. 
We express time in units of the inverse pumping rate $1/\Gamma$. 
We see that the transient time, i.e., the time needed to reach the steady state, is on the order of 4000 pumping cycles (a microsecond for the experiment). 
At very short time scales, the three-level atom shows a significant population imbalance; this accompanies a fast increase in the photon number in the cavity. 
In the steady state more than one hundred photons are present in the cavity, in agreement with the experimental estimates.
The photon distribution follows a binomial law (see Fig.~\ref{figtransphot}, inset), characteristic for correlated particles at zero temperature~\cite{NazarovBlanter}.
We next calculate the output spectrum Eq.~\eqref{spectrum}, as a function of the frequency $\omega$ and of the parameter $n_g$. 
The result is plotted in Fig.~\ref{fig2Dspectrum}. It presents a peak centered at the resonance.
Furthermore, in the experiment the presence of charge fluctuations widens the spectrum.
This broadening can be overcome by driving the cavity.
Figure~\ref{fig2Dspectrumdrive} represents the spectrum as a function of the driving strength. 
As $v$ increases, the initial Lorentzian is converted into a Dirac peak located at the driving frequency, thus emphasizing the lasing effect.
Charge noise can be strongly suppressed if a transmon qutrit is used instead of the standard CPB.

\section{Coupling to a two-level system}

The experimental spectrum in the $(n_g,\omega)$ plane shows an additional peak centered at $\omega_{10}\simeq1.5\,\omega_0$. 
Such a second peak is absent in our simulations, which is consistent with the fact that the coupling strength $g$ is below the threshold for two-photon masing.
Indeed, even when including the counter-rotating terms of the master equation that are neglected in the RWA and can lead to multi-photon processes, lasing effect occurs only at the resonance. 
Moreover, if the observed peak were due to two-photon processes, it should have been located at $\omega_{10}\simeq2\,\omega_0$. 
A possible source of additional resonances is the coupling to a two-level system (TLS).
We will consider two kinds of coupling, first a resonant coupling with the lasing transition of the CPB and second a dispersive coupling with both the CPB and the cavity.
To understand if it is possible to observe off-resonant lasing effect with a simple TLS we focus on the steady state photon number.

\begin{figure}[t]
   \centering
   \includegraphics[height=6.0cm]{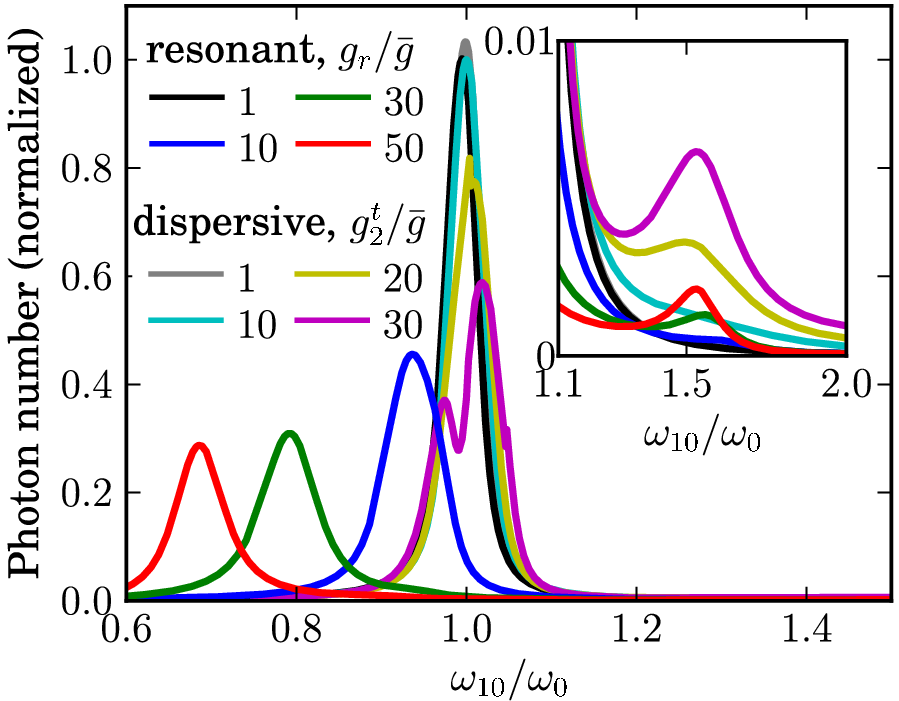}
   \caption{(Color online) Steady state photon number in the presence of a two-level system.
   The different lines correspond to different coupling strengths in the case of a resonant coupling and a dispersive one (see the legend, $\bar{g}/2\pi=44\,\mathrm{MHz}$).
   The inset is a zoom around the frequency $1.5\,\omega_0$, where the second peak appears.
   The damping $\kappa$ has been increased fivefold and the photon number is normalized by the number at the resonance without TLS.}
   \label{figTLS}
   \end{figure}

The Josephson junction of the CPB can be a source of fluctuators, resulting from the tunneling of a charge between two sites in the insulating layer~\cite{wellstood}.
The TLS is then composed of two states, ground state $\ket{g}$ and excited state $\ket{e}$, separated by an energy $\hbar\omega_f=\sqrt{E_f^2+4T^2}$, where $E_f$ is the energy difference between the two sites and $T$ is the tunneling strength.
The TLS is described by the Hamiltonian $H_f=\tfrac{1}{2}\hbar\omega_f\varsigma_z$, with $\varsigma_z=[\varsigma_+,\varsigma_-]$ where $\varsigma_+=|e\rangle\langle g|$ and $\varsigma_-=\varsigma_+^\dag$.
The tunneling charge position couples to the Cooper pair number, with an energy $\hbar g_r$, forming a four level system $\{\ket{0},\ket{1}\}\otimes\{\ket{g},\ket{e}\}$.
This system is furthermore coupled to the single electron state $\ket{2}$ with the incoherent pumping and to the cavity states through the Jaynes-Cummings Hamiltonian~\eqref{Heigen}.
While varying the gate voltage, different transitions of the four level system can become resonant with the cavity and induce lasing.
The interaction Hamiltonian turns out to comprise both a transverse and a longitudinal coupling, as well as frequency shifts,
   \begin{equation}
   H_r=\tfrac{1}{2}\hbar\nu_{10}\sigma_z+\tfrac{1}{2}\hbar\nu_f\varsigma_z+\hbar g_r^t(\sigma_-\varsigma_++\sigma_+\varsigma_-)+\hbar g_r^l\sigma_z\varsigma_z,
   \end{equation}
where $\nu_{10}=4g_rE_c(1-n_g)/\hbar\omega_{10}$, $\nu_f=g_rE_f(1-n_g)/\hbar\omega_f$, $g_r^t=-g_rE_JT/\hbar^2\omega_{10}\omega_f$, and $g_r^l=2g_r(1-n_g)E_cE_f/\hbar^2\omega_{10}\omega_f$.
The energy spectrum of the four level system $\{\ket{\psi_1},\ket{\psi_2},\ket{\psi_3},\ket{\psi_4}\}$ is given by $E_{1,4}=\mp\hbar(\varpi_{10}+\varpi_f)/2+\hbar g_l$ and $E_{2,3}=\mp\hbar\sqrt{(\varpi_{10}-\varpi_f)^2+4{g_r^t}^2}/2-\hbar g_l$, where $\varpi_{10,f}=\omega_{10,f}+\nu_{10,f}$.
The ground state is $\ket{\psi_1}=\ket{0,g}$, the highest state is $\ket{\psi_4}=\ket{1,e}$ and the central terms are obtained after rotating $\ket{0,e},\ket{1,g}$ by an angle $\alpha$ satisfying $\tan2\alpha=2g_t/(\varpi_f-\varpi_{10})$.
Finally, the coupling Hamiltonian with the cavity, obtained from Eq.~\eqref{Heigen}, reads
   \begin{equation}
   H_\mathrm{FLS}=\hbar g\left[\cos\alpha(S_{02}+S_{13})+\sin\alpha(S_{01}-S_{23})\right]a^\dag+\mathrm{H.c},
   \end{equation}
where $S_{ij}=|\psi_i\rangle\langle\psi_j|$.
A lasing effect thus occurs if the transition 2-0, 3-1, 1-0, or 3-2 is in resonance with the cavity and the corresponding coupling strength is large enough.
The photon number as a function of the frequency $\omega_{10}$ is plotted in Fig.~\ref{figTLS} for different values of the coupling strength $g_r$.
The frequency of the TLS is adjusted close to $\omega_0$ ($\pm10\%$) to observe the second resonance at $\omega_{10}=1.5\,\omega_0$, in the regime $E_f=T$.

A second resonance can also be produced by a fluctuator of frequency $\omega_f=1.5\,\omega_0$ coupled to both the Josephson atom and the cavity.
The cavity can then be indirectly excited when the transition $\omega_{10}$ approaches the resonance with the TLS, $\omega_{10}\simeq\omega_f$.
In this dispersive regime, we consider only the transverse coupling~\cite{lisenfeld},
   \begin{align}
   H_d=\hbar\left[g_1^t(\sigma_-+\sigma_+)+g_2^t(a+a^\dag)\right](\varsigma_++\varsigma_-),
   \label{Htls}
   \end{align}
where $g_{1/2}^t$ is the transverse coupling strength between the TLS and the Josephson atom/cavity.
The coupling $g_1^t$ has the same dependence on $n_g$ as $g$ while $g_2^t$ is $n_g$-independent.
The steady state photon number is plotted in Fig.~\ref{figTLS} as a function of the gate voltage through $\omega_{10}$ for $g_1^t=g$ and different values of $g_2^t/\bar{g}$.
The damping rate has been increased fivefold for numerical reasons, but this does not change the results qualitatively.

In both cases the effect of the TLS on the photon number is small even for strong couplings (see Fig.~\ref{figTLS}, inset).
An ultra-strong coupling between the fluctuator and the system, unrealistic for the experiment~\cite{Astafiev}, is needed to observe off-resonant lasing.
A simple TLS is thus unlikely to explain the second hot spot of the experiment.

\section{Semi-classical approximation}

  \begin{figure}[t]
   \centering
   \includegraphics[height=6.0cm]{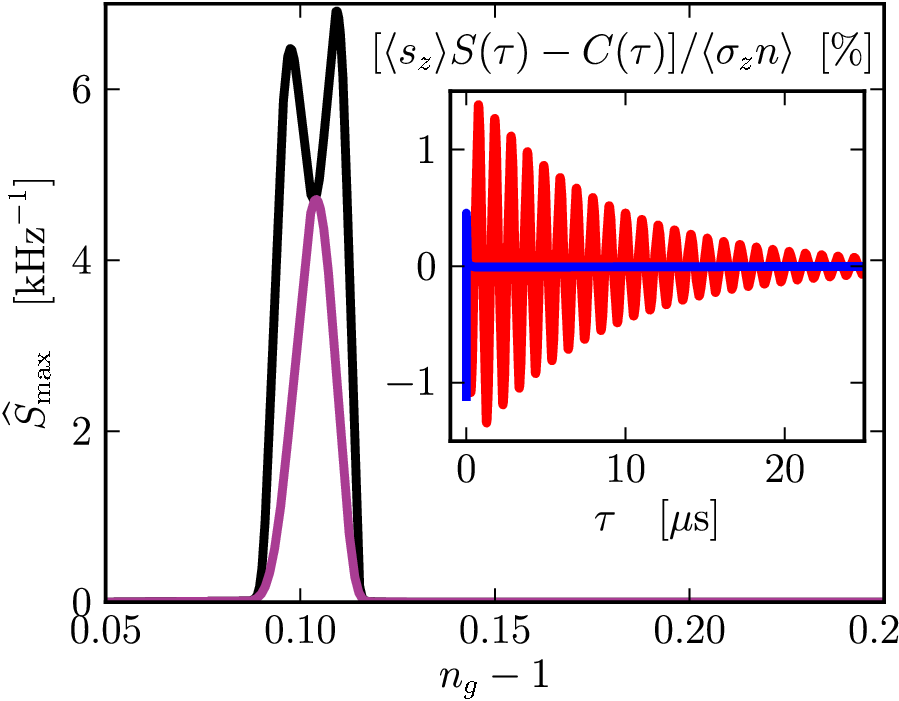}
   \caption{(Color online) Maximum value of the spectrum as a function of the reduced gate charge.
   The quantum solution is plotted in purple and the semi-classical one in black.
   Inset: Accuracy of the factorization Eq.~\eqref{factorization}.
   The time evolution of the real part of the normalized difference $[\langle s_z\rangle S(\tau)-C(\tau)]/\langle\sigma_zn\rangle$ in the rotating frame of the cavity is plotted at the resonance ($\omega_{10}=\omega_0$) in blue and at the second peak of $\widehat{S}_\mathrm{max}(n_g)$ ($\omega_{10}=1.06\,\omega_0$) in red.}
   \label{figSmax}
   \end{figure}

The results presented so far were obtained by numerically solving the Lindblad master equation. 
When the steady-state photon number is large, one can use the semi-classical approximation to get analytical results~\cite{brosco}. 
It consists of factorizing the operators pertaining to the three-level atom and to the cavity. 
The steady-state value of the photon number, for instance, is then obtained from a set of equations involving $\moy{\sigma_z}$, $\moy{\sigma_{11}}$, $\moy{\sigma_{01}a^\dag}$, and $\moy{\sigma_zn}$. 
The latter can be factorized in the semi-classical limit and the resulting solution is in good agreement with the numerical results.
At the level of the spectral function, the time derivative of $S(\tau)$ induces more complex correlators such as $C(\tau)=\moy{\sigma_z(\tau)a^\dag(\tau)a(0)}$.
Using the amplitude-phase representation of the operator $a$ and assuming that the correlation time of the phase fluctuations is much longer than that of the amplitude fluctuations, the factorization can be improved~\cite{brosco}
   \begin{equation}
   \moy{\sigma_z(\tau)a^\dag(\tau)a(0)}
   \simeq\frac{1}{2}\left(\moy{\sigma_z}+\frac{\moy{\sigma_z n}}{\moy{n}}\right)S(\tau).
   \label{factorization}
   \end{equation}
The set of differential equations leads to a Lorentzian spectrum of width
   $
   k=\kappa/2-2g^2(\Gamma+\gamma_{21})\moy{s_z}/[(\Gamma+\gamma_{21})^2+4\delta^2],
   $
and centered at the reduced frequency
   $
   \delta\omega=(4g^2\delta\moy{s_z})/[(\Gamma+\gamma_{21})^2+4\delta^2],
   $
where we note $\moy{s_z}=\frac{1}{2}\left(\moy{\sigma_z}+\moy{\sigma_z n}/\moy{n}\right)$
and $\delta=\omega_{10}-\omega_0$ is the detuning.
The maximum spectrum $\widehat{S}_\mathrm{max}$ with respect to the frequency $\omega$ is plotted as a function of the reduced charge gate $n_g$ in Fig.~\ref{figSmax} and compared to the Lorentzian solution in the semi-classical limit. 
The time evolution of the correlators in Eq.~\eqref{factorization} in the rotating frame of the cavity is shown in the inset, where the real part of the normalized difference $[\langle s_z\rangle S(\tau)-C(\tau)]/\langle\sigma_zn\rangle$ is plotted for two different values of the gate voltage.
At the resonance, the dynamics of the factorized correlator $\moy{s_z}S(\tau)$ is in good agreement with $C(\tau)$.
Off-resonance at $\omega_{10}=1.06\,\omega_0$, where the semi-classical spectrum exhibits the second peak, the difference oscillates at $\sim\omega_0/2\pi+1\,\mathrm{MHz}$, giving rise to a non-negligible contribution in the Fourier transform.
These comparisons reveal that the semi-classical treatment is not correct in the region close to the resonance where the correlations between the atom and the cavity cannot be neglected.
The resulting double peak structure thus appears to be an artifact of the factorization~\cite{andre}. 
Further improvement of the factorization, Eq.~\eqref{factorization}, is needed to describe the spectrum properly in the semi-classical limit.

\section{Conclusion}

In conclusion, the Lindblad master equation together with the quantum regression theorem are powerful tools to calculate quantum mechanically the time-evolution of the photon number and the output spectrum of the cavity. 
A comparison with the experimental results of Ref.~\onlinecite{Astafiev} gives access to the typical time-scales of the system. 
The calculation of the spectrum enables us to understand the experimental results and the effect of driving the cavity. 
It shows in particular that the presence of charge noise reduces the lasing effect. 
Considering the presence of a fluctuator in the system, we show that an ultra-strong coupling is needed to explain the second hot spot.
Finally, the fully quantum treatment for a three-level artificial atom, based on the density matrix of the whole system, allows to figure out the validity of the semi-classical approximations, which do not take into account all the correlations between the atom and the cavity.

\begin{acknowledgments}
The authors thank V.~Brosco, L.~I.~Glazman, A.~O.~Niskanen, and G.~Sch\"on as well as P.~Bertet, D.~Esteve, and D.~Vion from the Quantronics group of CEA-Saclay for useful discussions.
This work is supported by the ANR project QUANTJO.
\end{acknowledgments}

\end{document}